\documentclass[aps,a4paper,superscriptaddress,twocolumn,showpacs,preprintnumbers,amsmath,amssymb]{revtex4}
\usepackage{graphicx}
\usepackage{dcolumn}
\usepackage{color}
\usepackage{latexsym,amsfonts}
\usepackage{textcomp}
\usepackage{bm}

\usepackage{ulem}

\def\beq{\begin{equation}}
\def\eeq{\end{equation}}

\begin{document}

\title{The influence of the chameleon field potential on transition frequencies of\\ gravitationally bound quantum states of ultra-cold neutrons}

\author{A. N. Ivanov}\email{ivanov@kph.tuwien.ac.at}
\affiliation{Atominstitut, Technische Universit\"at Wien, Wiedner Hauptstra{\ss}e 8-10, A-1040 Wien, Austria}
\author{R. H\"ollwieser}
\affiliation{Atominstitut, Technische Universit\"at Wien, Wiedner Hauptstra{\ss}e 8-10, A-1040 Wien, Austria}
\author{T. Jenke}
\affiliation{Atominstitut, Technische Universit\"at Wien, Stadionallee 2, A-1020 Wien, Austria}
\author{M. Wellenzohen}
\affiliation{Atominstitut, Technische Universit\"at Wien, Wiedner Hauptstra{\ss}e 8-10, A-1040 Wien, Austria}
\author{H. Abele}\email{abele@ati.ac.at}
\affiliation{Atominstitut, Technische Universit\"at Wien, Stadionallee 2, A-1020 Wien, Austria}
\date{\today}

\begin{abstract}
We calculate the chameleon field potential for ultracold neutrons,
bouncing on top of one or between two neutron mirrors in the gravitational field of the Earth.
For the resulting non--linear equations of motion we give approximate analytical
solutions and compare them with exact numerical ones for which we
propose the analytical fit.
The obtained solutions may be used for the quantitative analysis of contributions of a
chameleon field to the transition frequencies of quantum states of ultra-cold neutrons bound in the gravitational field of the Earth.
\end{abstract}

\maketitle

\section{Introduction}
\label{sec:intro}
A chameleon field has been suggested to drive the current phase of
cosmic acceleration for a large class of scalar potentials. The
properties depend on the density of a matter to which it is
immersed. Because of this sensitivity on the environment, the field
was called chameleon, and it can couple directly to baryons with
gravitational strength on Earth but it would be essentially massless
on solar system scales \cite{Chameleon1,Waterhouse}.

An interaction of a chameleon field to an environment with a mass density
$\rho$ and its self--interaction are described by the effective
potential $V_{\rm eff}(\phi)$ \cite{Brax1}
\beq
\label{eq:1}
	V_{\rm eff}(\phi) = V(\phi) + \rho\,e^{\,\beta \phi/M},
\eeq
where $\beta$ is a coupling constant and
$M = 1/\sqrt{8\pi G_N} = 2.435\times 10^{18}\,{\rm GeV}$
is the Planck mass.

As has been pointed out by Ref.~\cite{Brax1}, ultra-cold neutrons,
bouncing in the gravitational field of the Earth above a mirror can be
a good laboratory for testing the existence of a chameleon
field. There has been found a solution of the equations of motion and
limits for the coupling constant $\beta$ can estimated from the
contribution of a chameleon field to the transition frequencies of the
quantum gravitational states of ultra-cold neutrons bouncing in the
gravitational field of the Earth.  Such resonant transitions between
quantum states of a neutron in the gravity potential have been measured
by the qBounce Collaboration \cite{Jenke1}.
In this experiment, ultra-cold neutrons form gravitationally bound quantum states
between a neutron mirror on bottom and another neutron mirror on top with a relative
distance $d$.
The upper neutron mirror has a rough surface and thereby filters out higher, unwanted states.
The whole system may be vibrated in vertical direction with frequency $\omega$ and amplitude $a$.
Behind this system, the neutron transmission is measured. It shows a significant reduction in count rate,
if the vibration frequency corresponds to the energy difference between two eigenstates.
With this method, the transition frequencies have been determined, which tests Newton's Law at short distances
and is therefore sensitive to any deviation like chameleon fields.
So far, such chameleon fields
have only been calculated in the infinitely large spatial region above one
mirror~\cite{Brax1}. For the experiment presented in~\cite{Jenke1}, it is necessary to use solutions of a
chameleon field localized between two mirrors for the estimate of the
coupling constant $\beta$.

As we show below the appearance of an addition mirror complicates the
problem of the calculation of a chameleon field substantially. As a
result we propose the analysis of all possible solutions of the
problem under consideration.

The paper is organized as follows: In section~\ref{sec:cham1mirror},
the chameleon field for ultra-cold neutrons, bouncing
in vacuum in the gravitational field of the Earth
above a horizontal mirror, is calculated.  We show that the solution
obtained in~\cite{Brax1} is exact for this problem.  In
sections~\ref{sec:cham2mirrors}, the calculation of the chameleon
field is adapted to the experimental setup used in \cite{Jenke1}. For
this purpose, we consider a symmetric geometry, where the horizontal
mirrors occupy the spatial regions $\textstyle z^2 > \frac{d^2}{4}$,
whereas ultra-cold neutrons bounce in the spatial region $\textstyle
z^2 \leq \frac{d^2}{4}$.  We obtain the solutions for a chameleon
field in the spatial region $\textstyle z^2 \le \frac{d^2}{4}$.  In
section~\ref{sec:cham-num} we solve the problem of a chameleon field
between two mirrors numerically without any approximation
and give an analytical fit of these solutions for
  arbitrary power $n$.

\section{The chameleon field coupled to ultra-cold neutrons bouncing above a mirror}
\label{sec:cham1mirror}
According to Ref.\cite{Brax1,Brax2}, the potential
$V(\phi)$ takes the form:
\beq
\label{eq:2}
	V(\phi) = \Lambda^4 + \frac{\Lambda^{4 + n}}{\phi^n}.
\eeq
The scale $\Lambda$ is chosen to be equal to $\Lambda = 2.4(1)\times
10^{- 12}\,{\rm GeV}$ \cite{Brax1}. Following \cite{Brax1} and keeping
only the linear term in the expansion of the exponential $\textstyle
e^{\,\beta \phi/M}$ in powers of $\phi$ one can show that the
effective potential has a minimum
\beq
\label{eq:3}
V'_{\rm
  eff}(\phi_{\rm min}) = V'(\phi_{\rm min}) + \rho\,\frac{\beta}{M} = 0,
\eeq
where $V'_{\rm eff}(\phi)$ is a derivative with respect to $\phi$, at
\beq
\label{eq:4}
\phi_{\rm min} = \Lambda\,\Big(\frac{nM\Lambda^3}{\beta \rho}\Big)^{1/(n+1)}.
\eeq
Quanta of a chameleon field are massive with mass defined by the
second derivative of the potential $V(\phi)$ \cite{Brax1}.  A
chameleon field satisfies an equation of motion \cite{Brax1}:
\beq
\label{eq:5}
\Box
\phi = - V'_{\rm eff}(\phi) = - V'(\phi) + V'(\phi_{\rm min}).
\eeq
As analysed by Brax {\it et al.} \cite{Brax1}, a neutron mirror with mass density $\rho_m$ being parallel to the $(x,y)$--plane at
$z = 0$, occupies a spatial region $z \leq 0$, whereas ultra-cold
neutrons bounce above in the gravity field in the spatial region $z \ge 0$ with vacuum density
$\rho_v \simeq 0$.\\
For such a geometry a chameleon field depends on the spatial
variable $z$ only and the equation of motion takes the form
\beq
	\frac{d^2\phi}{d z^2} =  V'(\phi) - V'(\phi_{\rm min}),
	\label{eq:6}
\eeq where $\phi_{\rm min} = \phi_m$ and $\phi_{\rm min} = \phi_v$ for
spatial regions $z \le 0$ and $z \ge 0$, respectively.  In our
analysis of a chameleon field for a system mirror--vacuum we follow
\cite{Brax1}, but -- in view of the following sections -- locate the
mirror at $\textstyle z = - \frac{d}{2}$.  In this case the mirror
occupies the spatial region $\textstyle z \le -\frac{d}{2}$, whereas
vacuum fills up the region $\textstyle z \ge - \frac{d}{2}$.

After the first integration the equation of motion Eq.~(\ref{eq:6})
reduces to the form
\beq
	\frac{1}{2}\Big(\frac{d\phi}{d z}\Big)^2 = C +
	V(\phi) - \phi V'(\phi_{\rm min}),
	\label{eq:7}
\eeq
where $C$ is a constant of integration. This constant we obtain from
the asymptotic conditions.
This constant we obtain
from the asymptotic conditions assuming that a chameleon field tends to
its minimum and a vanishing derivative at $z \to \pm \infty$ ~\cite{Brax1}:
\beq
		\lim_{z \to \pm\infty} \phi(z) = \phi_m\quad \lim_{z
		  \to \pm\infty} \frac{d\phi(z)}{d z} = 0.
		\label{eq:8}
\eeq
These conditions result in
\beq\label{eq:9}
C = - V(\phi_{\rm min}) + \phi_{\rm min} V'(\phi_{\rm
  min}).
\eeq
Substituting Eq.(\ref{eq:9}) into Eq.(\ref{eq:7}) we obtain
\beq
	\frac{1}{2}\Big(\frac{d\phi}{d z}\Big)^2 =
	V(\phi)  - V(\phi_{\rm min}) - (\phi - \phi_{\rm min}) V'(\phi_{\rm min}).
	\label{eq:10}
\eeq
Thus, in spatial regions $\textstyle z \le -\frac{d}{2}$ and
$\textstyle z \ge -\frac{d}{2}$ a chameleon field should obey the equations
\begin{eqnarray}\label{eq:11}
\frac{1}{2}\Big(\frac{d\phi}{d z}\Big)^2 &=& V(\phi) -
V(\phi_m) - (\phi - \phi_m) V'(\phi_m) = \nonumber\\
&=& \frac{\Lambda^{4+ n}}{\phi^n}\Big(1 - (n +
1)\,\frac{\phi^n}{\phi^n_m} + n\,\frac{\phi^{n + 1}}{\phi^{n +
    1}_m}\Big)
\end{eqnarray}
and
\begin{eqnarray}\label{eq:12}
\frac{1}{2}\Big(\frac{d\phi}{d z}\Big)^2 &=& V(\phi) -
V(\phi_v) - (\phi - \phi_v) V'(\phi_v) = \nonumber\\
&=& \frac{\Lambda^{4+ n}}{\phi^n}\Big(1 - (n +
1)\,\frac{\phi^n}{\phi^n_v} + n\,\frac{\phi^{n + 1}}{\phi^{n +
    1}_v}\Big),
\end{eqnarray}
respectively, where we have used the exact shape of the potential
$V(\phi)$, given by Eq.(\ref{eq:2}), and $\phi_m$ and $\phi_v$ are
given by Eq.(\ref{eq:4}) for $\rho = \rho_m$ and $\rho = \rho_v$,
respectively.

The solutions of Eq.(\ref{eq:11}) and Eq.(\ref{eq:12}) should
obey the boundary conditions
\begin{eqnarray}\label{eq:13}
 \phi(z)\Big|_{ z \to - \frac{d}{2}-} &=&
\phi(z)\Big|_{ z \to -\frac{d}{2}+} = \phi_d, \nonumber\\
\frac{d\phi(z)}{dz}\Big|_{ z \to -\frac{d}{2} -} &=&
\frac{d\phi(z)}{dz}\Big|_{ z \to -\frac{d}{2} +},
\end{eqnarray}
where $\phi_d = \phi(- d/2)$. Since ultra-cold neutrons are in the
spatial region $\textstyle z \ge - \frac{d}{2}$, we have to search a
solution of Eq.(\ref{eq:12}) only.  As regards Eq.(\ref{eq:11}),
it may be used only for the boundary conditions.

As a vacuum density $\rho_v \simeq 0$ leads to $\phi_v \to \infty$, we
take the r.h.s. of Eq.(\ref{eq:12}) in the limit $\phi_v \to
\infty$. This reduces Eq.(\ref{eq:12}) to the form
\beq\label{eq:14}
\frac{1}{2}\Big(\frac{d\phi}{d z}\Big)^2 =
\frac{\Lambda^{4+ n}}{\phi^n}.
\eeq
Since a chameleon field grows with $z \to \infty$, in the spatial
region $\textstyle z \ge - \frac{d}{2}$ a derivative of a chameleon
field is always positive. This gives a differential equation
\beq\label{eq:15}
\sqrt{2}\,\Lambda dz =
\Big(\frac{\phi}{\Lambda}\Big)^{\textstyle \frac{n}{2}}\frac{d\phi}{\Lambda}.
\eeq
The solution of this equation is given by the integral
\beq\label{eq:16}
\sqrt{2}\,\Lambda \Big(\frac{d}{2} + z\Big) =
\int^{\phi}_{\phi_d}\Big(\frac{\phi}{\Lambda}\Big)^{\textstyle
  \frac{n}{2}}\frac{d\phi}{\Lambda}.
\eeq
Integrating over $\phi$ and solving the obtained expression with
respect to $\phi(z)$ we obtain
\beq
		\phi(z) = \phi_d \Big(1 + \frac{n + 2}{\sqrt{2 n (n +
    1)}}\,\frac{1}{\lambda_d}\Big(\frac{d}{2} + z\Big)\Big)^{\textstyle \frac{2}{n + 2}},
    \label{eq:17}
\eeq
where $1/\lambda_d = \Lambda \sqrt{n(n + 1)(\Lambda/\phi_d)^{n +
    2}}$. At $\textstyle z = - \frac{d}{2}$ from a boundary condition
on the derivatives of a chameleon field
\beq\label{eq:18}
\frac{\Lambda^{4+ n}}{\phi^n_d}\Big(1 - (n +
1)\,\frac{\phi^n_d}{\phi^n_m} + n\,\frac{\phi^{n + 1}_d}{\phi^{n +
    1}_m}\Big) = \frac{\Lambda^{4+ n}}{\phi^n_d}
\eeq
we define $\phi_d$ in terms of $\phi_m$
\beq\label{eq:19}
\phi_d = \frac{n + 1}{n}\,\phi_m.
\eeq
This relation is valid for arbitrary value of the coupling constant
$\beta$.  Since the solution Eq.(\ref{eq:17}) obeys the asymptotic
conditions Eq.(\ref{eq:8}) at $z \to \infty$ and the boundary
conditions Eq.(\ref{eq:13}), it defines a chameleon field in the
spatial region $\textstyle z \ge - \frac{d}{2}$.

It is obvious that if a mirror with a density $\rho_m$ and vacuum with
a density $\rho_v \simeq 0$ occupy the spatial regions $\textstyle z
\ge + \frac{d}{2}$ and $\textstyle z \le + \frac{d}{2}$, respectively,
the solution of a chameleon field in the spatial region $\textstyle z
\le + \frac{d}{2}$ is
\beq\label{eq:20}
		\phi(z) = \phi_d \Big(1 + \frac{n + 2}{\sqrt{2 n (n +
    1)}}\,\frac{1}{\lambda_d}\Big(\frac{d}{2} - z\Big)\Big)^{\textstyle \frac{2}{n + 2}},
\eeq
where $\phi_d = \phi(+d/2) = \displaystyle \frac{n + 1}{n}\,\phi_m$.

The solutions Eq.(\ref{eq:16}) and Eq.(\ref{eq:20}) agree well
with the solution, obtained in \cite{Brax1}. Of course, the solution
for a chameleon field in the spatial region above a mirror,
i.e. $\textstyle z \ge - \frac{d}{2}$, is a meaningful from the point
of view of the contribution of a chameleon field to the transition
frequencies of the quantum gravitational states of ultra-cold neutrons
\cite{Brax1}. Following \cite{Brax1} we define a contribution of a
chameleon field to the gravitational potential of the Earth coupled to
ultra-cold neutrons above a mirror as
\beq\label{eq:21}
\Phi(z) = m g \left( \frac{d}{2} +  z \right) + \beta\,\frac{m}{M}\,\phi(z),
\eeq
where $m$ and $g$ are the neutron mass and the gravitational
acceleration, respectively. In comparison with \cite{Brax1} the mirror
is shifts down to $\textstyle z = - \frac{d}{2}$.

The calculation of the contribution of a chameleon field,
for example, to the transition frequencies of the
quantum gravitational states of ultra-cold neutrons is related to the
calculation of the matrix elements $\langle p|\phi(z)\rangle$, where
$|p\rangle$ is a low--lying quantum state of ultra-cold neutrons in
the gravitational field of the Earth \cite{QBouncer,AbeleQB1}. Due the
Airy functions, describing the gravitational states of ultra-cold
neutrons, the main contribution to matrix elements $\langle
p|\phi(z)|p\rangle$ comes from the region around $\textstyle
(\frac{d}{2} + z) \sim \ell_0 = (2 m^2g)^{-1/3} = 5.9\,{\rm \mu m}$,
which is a natural scale for neutron quantum states in the
gravitational field of the Earth \cite{QBouncer,AbeleQB1}. In this
case in the strong coupling limit $\beta \ge 10^5$ due to a relation
Eq.(\ref{eq:19}) and the experimental density of a mirror $\rho_m =
2.51\,{\rm g/cm^3}$ \cite{Jenke1} one may find that the contribution
of the first term in the parentheses of Eqs.(\ref{eq:17}) and
(\ref{eq:20}) can be neglected and the solutions for a chameleon field
take the form \beq\label{eq:22} \phi(z) = \Lambda \Big(\frac{n +
  2}{\sqrt{2}}\,\Lambda\Big(\frac{d}{2}\pm z\Big)\Big)^{\textstyle
  \frac{2}{n + 2}}, \eeq independent of the coupling constant $\beta$.
This agrees well with the result, obtained in \cite{Brax1}.

\section{The chameleon field coupled to ultra-cold neutrons
bouncing between two mirrors}
\label{sec:cham2mirrors}
Following the geometry used in \cite{Brax2} for the calculation of a
contribution of a chameleon field to the Casimir force induced
between two parallel plates perpendicular to the $z$--axis, we let
ultra-cold neutrons bounce in the region
$\textstyle z^2 \le \frac{d^2}4$ with a density $\rho_v$.
The regions $\textstyle z^2 \ge \frac{d^2}{4}$ are occupied by neutron mirrors with a density $\rho_m$ such as
$\rho_m \gg \rho_v$.

For such a geometry a chameleon field depends on the variable $z$
only. It is described by the equation of motion
\beq\label{eq:23}
\frac{d^2\phi}{d z^2} =  V'(\phi) - V'(\phi_{\rm min}).
\eeq
This equation is valid for a chameleon field in three spatial regions
$\textstyle z^2 \ge \frac{d^2}{4}$ and $\textstyle z^2 \le
\frac{d^2}{4}$. At $\textstyle z = \pm \frac{d}{2}$ a chameleon field
has to satisfy the standard boundary conditions
\begin{eqnarray}\label{eq:24}
 \phi(z)\Big|_{ z \to \pm \frac{d}{2}-} &=&
\phi(z)\Big|_{ z \to \pm \frac{d}{2}+} = \phi_d, \nonumber\\
\frac{d\phi(z)}{dz}\Big|_{ z \to \pm \frac{d}{2} -} &=&
\frac{d\phi(z)}{dz}\Big|_{ z \to \pm \frac{d}{2} +}.
\end{eqnarray}
In the spatial regions $\textstyle z^2 \ge \frac{d^2}{4}$ after the
first integration Eq.(\ref{eq:23}) reduces to the first order
differential equation (see Eq.(\ref{eq:11}))
\begin{eqnarray}\label{eq:25}
\frac{1}{2}\Big(\frac{d\phi}{dz}\Big)^2
  &=& V(\phi) - V(\phi_m) - (\phi - \phi_m)\, V'(\phi_m) = \nonumber\\
&=&\frac{\Lambda^{4+ n}}{\phi^n}\Big(1 - (n +
1)\,\frac{\phi^n}{\phi^n_m} + n\,\frac{\phi^{n + 1}}{\phi^{n +
    1}_m}\Big).
\end{eqnarray}
where the integration constants are defined from the asymptotic
conditions at $z \to \pm \infty$. Below we use Eq.(\ref{eq:25}) only
for the boundary conditions as we have done in the case of one mirror
in section~\ref{sec:cham1mirror}.

Before the integration of the equation of motion for a chameleon field
in the spatial region $\textstyle z^2 \le \frac{d^2}{4}$ we have to
accept that a chameleon field has a minimum at $\phi_v$, which is
caused by the properties of the effective potential $V_{\rm
  eff}(\phi)$ but not a spatial region of a location of a chameleon
field.  Since the region of a localisation of a chameleon field
$\textstyle z^2 \le \frac{d^2}{4}$ is finite, such a minimum can be
never reached inside.

After the first integration the equation of motion for a chameleon
field in $\textstyle z^2 \le \frac{d^2}{4}$ reduces to the first order
differential equation
\beq\label{eq:26}
\frac{1}{2}\Big(\frac{d\phi}{dz}\Big)^2 = \bar{C} + V(\phi) -
\phi\, V'(\phi_v),
\eeq
where $\bar{C}$ is a constant of integration. Since the asymptotic
regions $z \to \pm \infty$ are not reachable for a chameleon field in
the spatial region $\textstyle z^2 \le \frac{d^2}{4}$, a constant
$\bar{C}$ should be determined at other auxiliary conditions
\cite{Brax2}. In our approach it is convenient to redefine the
constant $\bar{C}$ as follows $\bar{C} = - \Lambda^4(1 - C)$.  Taking
into account that $\rho_v \simeq 0$ and $\phi_v \to \infty$ we
transcribe Eq.(\ref{eq:26}) into the form
\beq\label{eq:27}
\frac{1}{2}\Big(\frac{d\phi}{dz}\Big)^2 =
\Lambda^4\Big(C + \frac{\Lambda^n}{\phi^n}\Big).
\eeq
Since the second derivative of a chameleon field in the spatial region
$\textstyle z^2 \le \frac{d^2}{4}$ is always negative
\beq\label{eq:28}
\frac{d^2\phi}{dz^2} = V'(\phi) = - \frac{n \Lambda^{4
    + n}}{\phi^{n + 1}},
\eeq
a chameleon field can never reach a minimum between two mirrors.  This
agrees well with the assumption that a chameleon field reaches a
minimum only at $z \to \pm \infty$. Then, due to a symmetry of the
spatial regions invariant under a transformation $z
\longleftrightarrow - z$ a chameleon field as a scalar field should
satisfy the constraint $\phi(z) = \phi(-z)$. Below following
\cite{Brax1} we solve a problem of a chameleon field between two
mirrors assuming that a derivative of a chameleon field at $z = 0$ is
continuous.

\subsection{The chameleon field in the spatial region
 $\bf \textstyle z^2 \le \frac{d^2}{4}$ with continuous derivative at $\bf z =
  0$}
\label{subsec:1}

Due to a symmetry of a spatial region $\textstyle z^2 \le
\frac{d^2}{4}$ a requirement of a continuity of a derivative of a
chameleon field assumes that
\beq\label{eq:29}
\frac{d\phi}{dz}\Big|_{z = 0} = 0.
\eeq
Setting $z = 0$ in Eq.(\ref{eq:27}) and using the constraint
Eq.(\ref{eq:29}) we obtain the constant $C$
\beq\label{eq:30}
C = - \frac{\Lambda^{4 + n}}{\phi_0},
\eeq
where $\phi_0 = \phi(0)$. As a result Eq.(\ref{eq:27}) takes the form
\beq\label{eq:31}
\frac{1}{2}\Big(\frac{d\phi}{dz}\Big)^2 =
\Lambda^4\Big(\frac{\Lambda^n}{\phi^n} -
\frac{\Lambda^n}{\phi^n_0}\Big).
\eeq
A solution of this equation can be written in the following standard
form
\beq\label{eq:32}
\sqrt{2}\Lambda z = \pm \frac{1}{\Lambda^{\textstyle
    \frac{n + 2}{2}}}\int^{\phi}_{\phi_0}\frac{\varphi^{\textstyle
    \frac{n}{2}}d \varphi}{\displaystyle \sqrt{1 -
    \frac{\varphi^n}{\phi^n_0}}}.
\eeq
For the subsequent analysis of the solution of Eq.(\ref{eq:32}) we
propose to make the following changes of variables. First, making a
change of variables
\beq\label{eq:33}
 t = \sqrt{1 - \frac{\varphi^n}{\phi^n_0}}
\eeq
we transcribe Eq.(\ref{eq:32}) into the form
\beq\label{eq:34}
 \sqrt{\frac{n}{2(n + 1)}}\,\frac{z}{\lambda_0} = \mp
\int^{t(\phi)}_0\frac{dt}{\displaystyle (1 - t^2)^{\textstyle \frac{n
      - 2}{2 n}}},
\eeq
where $1/\lambda_0 = \Lambda \sqrt{n(n +
  1)}(\Lambda/\phi_0)^{(n+2)/2}$ and $t(\phi) = \sqrt{1
  -\phi^n/\phi^n_0}$.

The integral in the r.h.s. of Eq.(\ref{eq:34}) can be represented in
a simpler form. After the change $t = \sin\psi$ for $0 \le \psi \le
\frac{\pi}{2}$ we arrive at the equation
\beq\label{eq:35}
 \sqrt{\frac{n}{2(n + 1)}}\,\frac{z}{\lambda_0} = \mp
\int^{\psi(\phi)}_0\cos^{\textstyle \frac{2}{n}}\psi\,d\psi,
\eeq
where $\psi(\phi) = {\rm arcsin}\sqrt{1 - \phi^n/\phi^n_0}$.

The integral in the r.h.s. of Eq.(\ref{eq:34}) can be represented in
terms of the incomplete Beta function \cite{HMF72}. For this aim we
use a relation
\begin{eqnarray}\label{eq:36}
&&\int^{t(\phi)}_0\frac{dt}{\displaystyle (1 - t^2)^{\textstyle \frac{n
      - 2}{2 n}}}= \frac{1}{2}\,B\Big(t^2(\phi);\frac{1}{2},\frac{n
  + 2}{2n}\Big).
\end{eqnarray}
 In terms of the incomplete Beta function Eq.(\ref{eq:34}) reads
\begin{eqnarray}\label{eq:37}
&& \sqrt{\frac{n}{2(n + 1)}}\,\frac{z}{\lambda_0} =\mp
\frac{1}{2}\,B\Big(t^2(\phi);\frac{1}{2},\frac{n + 2}{2n}\Big).
\end{eqnarray}
 From the boundary conditions for a chameleon field at $\textstyle z =
 \mp \frac{d}{2}$ we obtain a relation between $\phi_d$ and $\phi_0$
\begin{eqnarray}\label{eq:38}
&& \sqrt{\frac{n}{2(n + 1)}}\,\frac{d}{\lambda_0} =
B\Big(1 - \frac{\phi^n_d}{\phi^n_0}; \frac{1}{2}, \frac{n +
  2}{2n}\Big).
\end{eqnarray}
Using the boundary conditions for the derivatives of a chameleon field
we relate $\phi_d$ to $\phi_m$ and $\phi_0$ as follows
\beq\label{eq:39}
 \phi_d = \frac{n + 1}{n}\,\phi_m -
\frac{\phi^{n+1}_m}{\phi^n_0}.
\eeq
Suppose that $\phi_0$ does not depend on $\beta$ in the strong
coupling limit $\beta \gg 1$. We confirm such a property of $\phi_0$
below for the exact solution obtained for $n = 2$. We also show that
for the experimental density $\rho_m = 2.51\,{\rm g/cm^3}$ of a mirror
and $d = 25.5\,{\rm \mu m}$ \cite{Jenke1} the independence of $\phi_0$
of the coupling constant $\beta$ starts for $\beta \ge 10^5$.  Since
in the strong coupling limit $\phi_m \gg \phi^{n+1}_m/\phi^n_0$,
$\phi_d$ is related to $\phi_m$ as
\beq\label{eq:40}
 \phi_d = \frac{n + 1}{n}\,\phi_m.
\eeq
Substituting Eq.(\ref{eq:40}) into Eq.(\ref{eq:38}) and taking
into account that in the strong coupling limit $\phi_0 \gg \phi_d$ we
arrive at the equation
\begin{eqnarray}\label{eq:41}
&& \sqrt{\frac{n}{2(n + 1)}}\,\frac{d}{\lambda_0} =
\sqrt{\pi}\,\frac{\displaystyle \Gamma\Big(\frac{n + 2}{2
    n}\Big)}{\displaystyle \Gamma\Big(\frac{n + 1}{n}\Big)}.
\end{eqnarray}
For the derivation of Eq.(\ref{eq:41}) we have used a relation
\cite{HMF72}
\beq\label{eq:42}
B\Big(1; \frac{1}{2}, \frac{n + 2}{2 n}\Big) =
\frac{\displaystyle \Gamma\Big(\frac{1}{2}\Big) \Gamma\Big(\frac{n +
    2}{2 n}\Big)}{\displaystyle \Gamma\Big(\frac{n + 1}{n}\Big)},
\eeq
where $\Gamma(a)$ is Euler's Gamma function and $\Gamma(1/2) =
\sqrt{\pi}$ \cite{HMF72}. Solving Eq.(\ref{eq:41}) with respect to
$\phi_0$ we obtain $\phi_0$ as a function of $\Lambda$, $d$ and $n$
only
\beq\label{eq:43}
\phi_0 =
\Lambda\Bigg(\frac{n}{\sqrt{2\pi}}\frac{\displaystyle
  \Gamma\Big(\frac{n + 1}{n}\Big)}{\displaystyle \Gamma\Big(\frac{n
    + 2}{2n}\Big)}\,\Lambda d\Bigg)^{\textstyle \frac{2}{n + 2}}.
\eeq
Thus, we may assert that in the spatial region $\textstyle z^2 \le
\frac{d^2}{4}$ in the strong coupling limit $\beta \ge 10^5$, being
valid for the experimental density of a mirror $\rho_m = 2.51\,{\rm
  g/cm^3}$ \cite{Jenke1}and $d = 25.5\,{\rm \mu m}$ , any solution of
a chameleon field with a continuous derivative at $z = 0$ should
depend on two parameters $\textstyle \phi_d = \frac{n + 1}{n}\,\phi_m$
and $\phi_0$, determined by Eq.(\ref{eq:4}) for $\rho = \rho_m$ and
Eq.(\ref{eq:43}), respectively. This assertion we prove below for
the exactly solvable case $n = 2$.

\subsection{The chameleon field in the spatial region
 $\bf \textstyle z^2 \le \frac{d^2}{4}$ with continuous derivative at $\bf z =
  0$. Exact solution for $\bf n = 2$}
\label{subsec:2}

As it is seen from Eq.(\ref{eq:32}) for $n = 2$ the integral over
$\varphi$ can be calculated in therms of elementary functions.
Setting $n = 2$ and integrating over $\varphi$ we obtain
\beq\label{eq:44}
\sqrt{2}\Lambda z = \mp
\frac{\phi^2_0}{\Lambda^2}\sqrt{1 - \frac{\phi^2}{\phi^2_0}}.
\eeq
This defines  $\phi$ as a function of $z$
\beq\label{eq:45}
\phi(z) = \phi_0\Big(1 -
\frac{1}{3}\,\frac{z^2}{\lambda^2_0}\Big)^{1/2},
\eeq
where $1/\lambda_0 = \Lambda \sqrt{6}(\Lambda/\phi_0)^2$. At the next
step we have to express the parameter $\phi_0$ in terms of $\phi_d =
\phi(\pm d/2)$. Setting $\textstyle z = \pm \frac{d}{2}$ we obtain
\beq\label{eq:46}
\frac{\phi^2_d}{\phi^2_0} = 1 -
\frac{d^2}{12 \lambda^2_d}\,\frac{\phi^4_d}{\phi^4_0},
\eeq
where $1/\lambda_d = \Lambda \sqrt{6}(\Lambda/\phi_d)^2$.  The
solution of this algebraical equation with respect to $\phi^2_0$ s
\beq\label{eq:47}
\frac{\phi^2_d}{\phi^2_0} =
\frac{6\lambda^2_d}{d^2}\Big(\sqrt{1 + \frac{d^2}{3\lambda^2_d}} -
1\Big) = \frac{2}{\displaystyle 1 + \sqrt{1 +
    \frac{d^2}{3\lambda^2_d}}}.
\eeq
Thus, the solution for a chameleon field, defined for $n = 2$, is
\beq\label{eq:48}
\phi(z) = \phi_d\sqrt{1 +
  \frac{1}{3}\,\frac{1}{\Lambda^2_d}\Big(\frac{d^2}{4} - z^2\Big)},
\eeq
where $\Lambda_d$ is defined by
\beq\label{eq:49}
\Lambda_d = \lambda_d\sqrt{ \frac{\displaystyle 1 +
    \sqrt{1 + \frac{d^2}{3\lambda^2_d}}}{2}} =
\lambda_d\,\frac{\phi_0}{\phi_d}.
\eeq
Using the relation
\begin{eqnarray}\label{eq:50}
&& \frac{1}{3}\,\frac{\phi^4_d}{\Lambda^2_d}\,\Big(1 +
\frac{1}{12}\,\frac{d^2}{\Lambda^2_d}\Big) =
\frac{1}{3}\,\frac{\phi^4_d}{\lambda^2_d}\frac{2}{\displaystyle 1 +
  \sqrt{1 + \frac{d^2}{3\lambda^2_d}}}\nonumber\\
&&\times \Bigg(1 + \frac{d^2}{12\lambda^2_d}\,
\frac{2}{\displaystyle 1 + \sqrt{1 + \frac{d^2}{3\lambda^2_d}}}\Bigg)
= \nonumber\\
&&=
\frac{1}{3}\,\frac{\phi^4_d}{\lambda^2_d}\frac{2}{\displaystyle
  \Bigg(1 + \sqrt{1 + \frac{d^2}{3\lambda^2_d}}\,\Bigg)^2} \Bigg(1 +
\sqrt{1 + \frac{d^2}{3\lambda^2_d}} + \frac{d^2}{6\lambda^2_d}\Bigg) =
\nonumber\\
&&= \frac{1}{3}\,\frac{\phi^4_d}{\lambda^2_d}=
\frac{1}{3}\,\times \,\phi^4_d\,\times \,6\,\times
\,\frac{\Lambda^6}{\phi^4_d} = 2\times \,\Lambda^6
\end{eqnarray}
one can show that the solution Eq.(\ref{eq:48}) satisfies
Eq.(\ref{eq:28}) at $n = 2$.

From the boundary conditions for the first derivatives of a chameleon
field we may define the parameter $\phi_d$ in terms of $\phi_m$
\beq\label{eq:51}
\phi_d = \frac{3}{2}\,\phi_m -
\frac{1}{2}\frac{\phi^3_m}{\phi^2_0}.
\eeq
If $\phi_0$ does not depend on the coupling constant $\beta$ the
second term in the r.h.s. of Eq.(\ref{eq:51}), taken in the strong
coupling limit $\beta \ge 10^5$, should be smaller compared with the
first one. Neglecting the contribution of the second term we obtain
that in the strong coupling limit $\phi_d$ can be approximated by
$\textstyle \phi_d = \frac{3}{2}\,\phi_m$. This agrees well with
Eq.(\ref{eq:40}) at $n = 2$. Since for $n = 2$ the problem is
exactly solvable, we may check such a supposition by using the exact
relation between $\phi_d$ and $\phi_0$, given by Eq.(\ref{eq:47}),
and analysing the solution of Eq.(\ref{eq:51} numerically.

For this aim we transcribe Eq.(\ref{eq:51}) into the form
\beq\label{eq:52}
2\frac{\phi^3_d}{\phi^3_m} - 3
\frac{\phi^2_d}{\phi^2_m} + \frac{2}{\displaystyle 1 + \sqrt{1 +
    \frac{d^2}{3\lambda^2_d}}} = 0,
\eeq
where we have used Eq.(\ref{eq:47}).  Denoting $X = \phi_d/\phi_m$
we rewrite Eq.(\ref{eq:52}) as follows
\beq\label{eq:53}
f(X) = X^3 - \frac{3}{2} X^2 + \frac{1}{\displaystyle 1 +
  \sqrt{1 + \frac{d^2}{3\lambda^2_m}\,\frac{1}{X^4}}} = 0,
\eeq
where $\lambda_m$ is defined by
\beq\label{eq:54}
\frac{1}{\lambda_m} = \Lambda
\sqrt{6}\Big(\frac{\Lambda}{\phi_m}\Big)^2 = \Lambda
\sqrt{6}\Big(\frac{\beta \rho_m}{2 M\Lambda^3}\Big)^{2/3}.
\eeq
For numerical analysis we use $\rho_m = 2.51\,{\rm g/cm^3}$ and $d =
25.5\,{\rm \mu m}$ \cite{Jenke1}. One can show that the function
$f(X)$ has a real root, which for $\beta \ge 10^5$ is practically
equal to $X = 3/2$. This confirms our supposition that the parameter
$\phi_0$ does not depend on the coupling constant $\beta$ in the
strong coupling limit.

In the strong coupling limit $\beta \ge 10^5$, where $\phi_d =
(3/2)\,\phi_m$, the ratio $\phi^2_d/\phi^2_0$ is given by
\beq\label{eq:55}
\frac{\phi^2_d}{\phi^2_0} =
2\sqrt{3}\,\frac{\lambda_d}{d} = \frac{2\sqrt{3}}{d}\,\frac{1}{\Lambda
  \sqrt{6}}\frac{\phi^2_d}{\Lambda^2} = \frac{\sqrt{2}}{\Lambda
  d}\,\frac{\phi^2_d}{\Lambda^2}.
\eeq
Since $\phi^2_d$ is cancelled, we obtain $\phi_0$ as a function of
$\Lambda$ and $d$
\beq\label{eq:56}
\phi_0 = \frac{\Lambda}{2^{1/4}}\sqrt{\Lambda d}.
\eeq
This confirms our assertion about the independence of $\phi_0$ of the
coupling constant $\beta$ in the strong coupling limit and,
correspondingly, Eq.(\ref{eq:43}) for $n = 2$.  The scale
$\lambda_0$ is equal to
\beq\label{eq:57}
\lambda_0 = \frac{1}{\Lambda
  \sqrt{6}}\,\frac{\phi^2_0}{\Lambda^2} = \frac{1}{\Lambda
  \sqrt{6}}\,\frac{1}{\Lambda^2}\,\frac{1}{\sqrt{2}}\,\Lambda^2(\Lambda
d) = \frac{d}{2\sqrt{3}}.
\eeq
Thus, in the strong coupling limit the solution Eq.(\ref{eq:48}) can
be transcribed into the form
\begin{eqnarray}\label{eq:58}
\phi(z) &=& \frac{\Lambda}{2^{1/4}}\sqrt{\Lambda
  d}\,\Big(1 - 4\,\frac{z^2}{d^2}\Big)^{1/2} =\nonumber\\
&=& 2^{3/4}\sqrt{\frac{\Lambda}{d}}\,
\,\Big(\Lambda^2\Big(\frac{d^2}{4} - z^2\Big)\Big)^{1/2}.
\end{eqnarray}
In the vicinity of $\textstyle z \simeq \mp \frac{d}{2}$, where the
contribution of the mirror, localised at $\textstyle z = \pm
\frac{d}{2}$, can be neglected and the problem under consideration
reduces to the problem of a chameleon field coupled to ultra-cold
neutrons, bouncing in the gravitational field of the Earth above a
mirror, the solution Eq.(\ref{eq:58}) takes the form
\beq\label{eq:59}
\phi(z) = 2^{3/4}\Lambda \,\Big(\Lambda\Big(\frac{d}{2}
\pm z\Big)\Big)^{1/2}.
\eeq
It agrees well with the solution Eq.(\ref{eq:22}), taken in the
strong coupling limit.

As we have shown the exact solution of a chameleon field in the
spatial region $\textstyle z^2 \le \frac{d^2}{4}$ , carried out for $n
= 2$, confirms fully our suppositions, concerning a dependence of the
parameters $\phi_d$ and $\phi_0$ on the coupling constant $\beta$ in
the strong coupling limit.

The solution Eq.(\ref{eq:58}) can be applied to the analysis of the
contribution of a chameleon field to the transition frequencies of the
quantum gravitational states of ultra-cold neutrons bouncing in the
gravitational field between two mirrors. Following \cite{Brax1}, the
gravitational potential, corrected by the contribution of a chameleon
field, takes the form Eq.(\ref{eq:21}).

As we have already pointed out that due to the Airy wave functions the
solution of a chameleon field is localised in the vicinity of
$\textstyle (\frac{d}{2} + z) \sim \ell_0 = (2 m^2g)^{-1/3} =
5.9\,{\rm \mu m}$, which is a natural scale for neutron quantum states
in the gravitational field of the Earth
\cite{QBouncer,AbeleQB1}. Indeed, since a contribution of a chameleon
field to the gravitational potential of the interaction with ultra-cold
neutrons to the matrix elements $\langle p|\phi| p \rangle$ for
low--lying gravitational states $|p\rangle = |1\rangle$ and
$|3\rangle$ is localised around $\textstyle (\frac{d}{2} + z) \sim
\ell_0$ due to the Airy functions \cite{QBouncer,AbeleQB1}, the
solution Eq.(\ref{eq:58}), obtained in the strong coupling limit
$\beta \ge 10^5$, should be valid around $\textstyle (\frac{d}{2} + z)
\sim \ell_0$. This agrees also well with \cite{Brax1}.

\subsection{The chameleon field in the spatial region
 $\bf \textstyle z^2 \le \frac{d^2}{4}$ with continuous derivative
  at $\bf z = 0$. Approximate solution of non--linear equation for
  arbitrary $\bf n$}
\label{subsec:3}

Now we may proceed to solving Eq.(\ref{eq:34}) for arbitrary $n$. The
only approximation, which may lead to an analytical solution of the
problem, is $t \ll 1$.  This allows to get an agreement with the exact
solution at $n = 2$. Such an approximation gives us
the so--cold \textit{low--contrast solution}~\cite{Waterhouse}.
Neglecting in Eq.(\ref{eq:34}) the term $t^2$ with respect to a unity
and integrating over $t$ we arrive at the equation \beq\label{eq:60}
\sqrt{2}\,\Lambda z = \mp
\frac{2}{n}\,\Big(\frac{\phi_0}{\Lambda}\Big)^{\textstyle \frac{n +
    2}{2}} \sqrt{1 - \frac{\phi^n}{\phi^n_0}}.  \eeq Solving this
equation with respect to $\phi$ we obtain a solution for a chameleon
field \beq\label{eq:61} \phi(z) = \phi_0 \Big(1 - \frac{n}{2(n +
  1)}\,\frac{z^2}{\lambda^2_0}\Big)^{1/n}, \eeq where $1/\lambda_0 =
\Lambda \sqrt{n(n + 1)(\Lambda/\phi_0)^{n + 2}}$. At $n = 2$
Eq.(\ref{eq:61}) reduces to Eq.(\ref{eq:45}).

Now we have to replace the parameter $\phi_0$ by $\phi_d$. Setting
$\textstyle z = \pm \frac{d}{2}$ we obtain
\beq\label{eq:62}
\frac{\phi^n_d}{\phi^n_0} = 1 - \frac{n}{8(n +
  1)}\frac{d^2}{ \lambda^2_d}\,\frac{\phi^{n+2}_d}{\phi^{n+2}_0},
\eeq
where $1/\lambda_d = \Lambda \sqrt{n(n + 1)(\Lambda/\phi_d)^{n +
    2}}$. For the subsequent calculations it is convenient to rewrite
Eq.(\ref{eq:62}) as follows
\beq\label{eq:63}
\frac{\phi^n_d}{\phi^n_0} = \frac{1}{\displaystyle 1 +
  \frac{n}{8(n + 1)}\frac{d^2}{
    \lambda^2_d}\,\frac{\phi^2_d}{\phi^2_0}}.
\eeq
Using Eq.(\ref{eq:63}) we transcribe Eq.(\ref{eq:61}) into the form
\begin{eqnarray}\label{eq:64}
&& \phi(z) = \phi_0 \Big(1 - \frac{n}{2(n +
  1)}\,\frac{z^2}{\lambda^2_0}\Big)^{1/n} =\nonumber\\
&& = \phi_0 \Big(1 - \frac{n}{2(n +
  1)}\,\frac{z^2}{\lambda^2_d}\frac{\phi^{n+2}_d}{\phi^{n +
    2}_0}\Big)^{1/n} =\nonumber\\
&&= \phi_0 \Bigg(1 - \frac{n}{2(n +
  1)}\,\frac{z^2}{\lambda^2_d}\frac{\phi^2_d}{\phi^2_0}\frac{1}{\displaystyle
  1 + \frac{n}{8(n + 1)}\frac{d^2}{
    \lambda^2_d}\,\frac{\phi^2_d}{\phi^2_0}}\Bigg)^{1/n} =\nonumber\\
&&= \frac{\phi_0}{\displaystyle \Big(1 + \frac{n}{8(n +
    1)}\frac{d^2}{\lambda^2_d}\frac{\phi^2_d}{\phi^2_0}\Big)^{1/n}}\nonumber\\
&&\times \Bigg(1 + \frac{n}{2(n +
  1)}\frac{1}{\lambda^2_d}\Big(\frac{d^2}{4} -
z^2\Big)\,\frac{\phi^2_d}{\phi^2_0}\Bigg)^{1/n}=\nonumber\\
&&= \phi_d \Bigg(1 + \frac{n}{2(n +
  1)}\frac{1}{\Lambda^2_d}\Big(\frac{d^2}{4} -
z^2\Big)\Bigg)^{1/n},
\end{eqnarray}
where we have denoted $\Lambda_d = \lambda_d(\phi_0/\phi_d)$. Thus, an
approximate solution for a chameleon field for an arbitrary $n$ is
\beq\label{eq:65}
\phi(z) = \phi_d \Bigg(1 + \frac{n}{2(n +
  1)}\frac{1}{\Lambda^2_d}\Big(\frac{d^2}{4} - z^2\Big)\Bigg)^{1/n}.
\eeq
For $n = 2$ the function Eq.(\ref{eq:61}) reduces to the function
Eq.(\ref{eq:48}).

From Eq.(\ref{eq:62}) in the strong coupling limit we define
$\phi_0$ by the expression
\beq\label{eq:66}
\phi_0 = \Lambda\Big(\frac{n}{2\sqrt{2}}\,\Lambda
d\Big)^{\textstyle \frac{2}{n + 2}}.
\eeq
For $1 \le n \le 10$ such an expression reproduces the exact result
Eq.(\ref{eq:43}) with an accuracy better than $6.3\,\%$. Thus, we
may assert that the property of $\phi_0$ to be independent of the
coupling constant $\beta$ in the strong coupling limit is a general
property, which does not depend on the approximation, but an exact
dependence of $\phi_0$ on the parameters $\Lambda$. $d$ and $n$
depends, of course, on it.

In order to show that the function Eq.(\ref{eq:65}) can be used as a
solution for a chameleon field in the spatial region $\textstyle z^2
\le \frac{d^2}{4}$ and satisfies the equation of motion for a chameleon
field, we have to derive the equation of motion Eq.(\ref{eq:28}) at
the same approximation, which we have used for the derivation of
Eq.(\ref{eq:65}).  For this aim we have to derive the equation of
motion for $t(\phi)$. The first derivative of $\phi$ is equal to
\beq\label{eq:67}
\frac{d\phi}{dz} = - \frac{2}{n}\,\phi_0\,t(1 - t^2)^{\textstyle \frac{1}{n} -
  1}\frac{dt}{dz}
\eeq
and the second one is
\begin{eqnarray}\label{eq:68}
\frac{d^2\phi}{dz^2} &=& - \frac{2}{n}\,\phi_0\,t(1 - t^2)^{\textstyle\frac{1}{n}
  - 1}\frac{d^2t}{dz^2}\nonumber\\
&&- \frac{2}{n}\,\phi_0\,(1 - t^2)^{\textstyle\frac{1}{n} -
  1}\Big(\frac{dt}{dz}\Big)^2\nonumber\\
&& - \frac{2}{n}\frac{2(n - 1)}{n}\,\phi_0\,(1 -
t^2)^{\textstyle\frac{1}{n} - 2}\Big(\frac{dt}{dz}\Big)^2.
\end{eqnarray}
The r.h.s. of Eq.(\ref{eq:28}) can be transcribed into the form
\begin{eqnarray}\label{eq:69}
&&-\frac{n \Lambda^{n + 4}}{\phi^{n + 1}} = - \frac{n
  \Lambda^{n + 4}}{\phi^{n + 1}_0}\frac{1}{\displaystyle (1 - t^2)^{\textstyle\frac{n + 1}{n}}}.
\end{eqnarray}
Thus, the equation of motion Eq.(\ref{eq:28}), rewritten for
$t(\phi)$, takes the form
\begin{eqnarray}\label{eq:70}
&&t\frac{d^2t}{dz^2} + \Big(\frac{dt}{dz}\Big)^2 +
\frac{2(n - 1)}{n}\,\frac{t^2}{1 - t^2}\,\Big(\frac{dt}{dz}\Big)^2 =\nonumber\\
&&= \frac{n^2}{2}\frac{ \Lambda^{n + 4}}{\phi^{n +
    1}_0}\frac{1}{\displaystyle (1 - t^2)^{\textstyle\frac{2}{n}}}.
\end{eqnarray}
Neglecting the contributions of the terms of order $O(t^2)$ we reduce
Eq.(\ref{eq:70}) to the form
\begin{eqnarray}\label{eq:71}
&&t\frac{d^2t}{dz^2} + \Big(\frac{dt}{dz}\Big)^2 =
\frac{n^2}{2}\frac{ \Lambda^{n + 4}}{\phi^{n +
    2}_0}.
\end{eqnarray}
In our approximation the solution of Eq.(\ref{eq:34}) is
\begin{eqnarray}\label{eq:72}
&&t = \pm
\frac{n}{\sqrt{2}}\,\frac{\Lambda^{\textstyle\frac{{n +
        4}}{2}}}{\phi^{\textstyle\frac{n + 2}{2}}_0}\,z.
\end{eqnarray}
Substituting Eq.(\ref{eq:72}) into Eq.(\ref{eq:72}) we satisfy the
approximate equation of motion. This means that we may use the
solution Eq.(\ref{eq:65}) for a chameleon field in the spatial
region $\textstyle z^2 \le \frac{d^2}{4}$.

Since we are interested in the solution for a chameleon field in the
strong coupling limit, using the relations $\textstyle \phi_d =
\frac{n + 1}{n }\,\phi_m$ and Eq.(\ref{eq:66}) we reduce the
solution Eq.(\ref{eq:65}) to the form
\begin{eqnarray}\label{eq:73}
\phi(z) &=& \Lambda\,2^{\textstyle
  \frac{n-4}{n(n+2)}}n^{\textstyle \frac{2}{n + 2}}(\Lambda
d)^{\textstyle -\frac{4}{n(n + 2)}}\nonumber\\
&&\times\,\Big(\Lambda^2\Big(\frac{d^2}{4} -
z^2\Big)\Big)^{\textstyle \frac{1}{n}}.
\end{eqnarray}
We would like to remind that the solution Eq.(\ref{eq:73}) is valid
for $\beta \ge 10^5$ if the mirror density is equal to $\rho_m =
2.51\,{\rm g/cm^3}$ and $d = 25.5\,{\rm g/cm^3}$ \cite{Jenke1}.  For
$n = 2$ it coincides with Eq.(\ref{eq:58}). In the vicinity of the
mirrors $\textstyle z = \mp \frac{d}{2}$, the solution
Eq.(\ref{eq:73}) takes the form
\beq\label{eq:74}
\phi(z) = \Lambda\,2^{\textstyle \frac{n-4}{n(n+2)}}n^{\textstyle
  \frac{2}{n + 2}}(\Lambda d)^{\textstyle \frac{n - 2}{n +
    2}}\Big(\Lambda\Big(\frac{d}{2} \pm z\Big)\Big)^{\textstyle
  \frac{1}{n}}.
\eeq
One may see that the solution Eq.(\ref{eq:74}) reproduces the
solution for a chameleon field above (below) a mirror (see
Eq.(\ref{eq:22})) only for $n = 2$.

We would like to note that it is obvious that even if the solution for
a chameleon field with $n = 2$ and arbitrary coupling constant $\beta$
does not reproduce in the vicinity of a mirror the solution for a
chameleon field above one mirror. This means that fact that the
solution Eq.(\ref{eq:73}) does not reproduce in the vicinity of a
mirror the solution for a chameleon field above a mirror (see
Eq.(\ref{eq:22}) should not be evaluated as a strong argument
against to use such a solution for the problem under consideration.

That is why in section~\ref{sec:cham-num} we apply the solution Eq.(\ref{eq:73}) to
the estimate of the coupling constant $\beta$ from the contribution of
a chameleon field to the transition frequencies of quantum
gravitational states of ultra-cold neutrons, bouncing in the
gravitational field of the Earth between two mirrors \cite{Jenke1}.

\subsection{The chameleon field in the spatial region
 $\bf \textstyle z^2 \le \frac{d^2}{4}$ with continuous derivative at $\bf z =
  0$. Solution of the linearised equation of motion for arbitrary $\bf n$}
\label{subsec:4}

In order to complete the analysis of solutions for a chameleon field,
possessing a continuous derivative at $z = 0$, we have to consider the
linearised equations of motion. In the spatial region $\textstyle z^2
\le \frac{d^2}{4}$ the linearised Eq.(\ref{eq:28}) takes the form
\beq\label{eq:75}
\frac{d^2\phi}{dz^2} = - n\frac{\Lambda^{4 +
    n}}{\phi^{n + 1}_0} -n(n+1)\frac{\Lambda^{4 + n}}{\phi^{n +
    2}_0}(\phi_0 - \phi).
\eeq
Denoting $\phi_0 - \phi = \varphi$ we define the following linearised
equation for a chameleon field in the spatial region $\textstyle z^2
\le \frac{d^2}{4}$
\beq\label{eq:76}
\frac{d^2\varphi}{dz^2} - \frac{\varphi}{\lambda^2_0}=
\frac{n \Lambda^{4 + n}}{\phi^{n + 1}_0},
\eeq
where $1/\lambda^2_0 = n(n+1)\Lambda^{n+4}/\phi^{n+2}_0$. As $\phi(z )
= \phi(-z)$ the solution of Eq.(\ref{eq:76}) is
\beq\label{eq:77}
\varphi(z) = - \lambda^2_0\,\frac{n \Lambda^{4 +
    n}}{\phi^{n + 1}_0} + A\,\cosh\Big(\frac{z}{\lambda_0}\Big),
\eeq
where $A$ is a constant of integration.  Since $\varphi(0) = 0$, we
get
\begin{eqnarray}\label{eq:78}
&&\varphi(z) = - \lambda^2_0\,\frac{\Lambda^{4 +
    n}}{\phi^{n + 1}_0}\Big(1 -
\cosh\Big(\frac{z}{\lambda_0}\Big)\Big) = \lambda^2_0\,\frac{2
  n\Lambda^{4 + n}}{\phi^{n + 1}_0}\nonumber\\
&&\times\, \sinh^2\Big(\frac{z}{2\lambda_0}\Big) =
\phi_0\,\frac{2}{n + 1}\, \sinh^2\Big(\frac{z}{2\lambda_0}\Big).
\end{eqnarray}
This gives a chameleon field $\phi(z)$ equal to
\beq\label{eq:79}
\phi(z) = \phi_0\Big(1 - \frac{2}{n +
  1}\sinh^2\Big(\frac{z}{2\lambda_0}\Big)\Big).
\eeq
This solution agrees well with that obtained in \cite{Brax2} (see
Eq.(\ref{eq:20}) of \cite{Brax2} at $V'_b = 0$).

Since we are interesting in the solution for a chameleon field in the
strong coupling limit, the solution Eq.(\ref{eq:79}) may be valid in
the strong coupling limit if it satisfies the following
constraint. Indeed, in the strong coupling limit $\phi_d$ is
proportional to $\phi_m$, i.e. $\textstyle \phi_d =\frac{n +
  1}{n}\,\phi_m$, and commensurable with zero. This implies that the
solution Eq.(\ref{eq:79}) should vanish at $\textstyle z = \pm
\frac{d}{2}$ (see Eq.(\ref{eq:58}) and Eq.(\ref{eq:73})). Setting
$\textstyle z = \pm \frac{d}{2}$ we obtain a constraint on $\phi_0$
\begin{eqnarray}\label{eq:80}
&&\phi_0 = \Lambda \Bigg(\frac{\sqrt{n(n +
    1)}}{4}\,\frac{\Lambda d}{\displaystyle {\ell n}\Big(\sqrt{\frac{n
      + 1}{2}}+\sqrt{\frac{n + 3}{2}}\;\Big) }\Bigg)^{\textstyle
  \frac{2}{n + 2}}.\nonumber\\
&&
\end{eqnarray}
One can show that for $1 \le n \le 10$ the parameter $\phi_0$, defined
by Eq.(\ref{eq:80}), fits the exact expression Eq.(\ref{eq:43})
with an accuracy better than $8.7\,\%$.

The solution Eq.(\ref{eq:80}) with $\phi_0$, defined by
Eq.(\ref{eq:80}), can be also applied to the estimate of the
coupling constant $\beta$ from the contribution of a chameleon field
to the transition frequencies of quantum gravitational states of
ultra-cold neutrons, bouncing in the gravitational field of the Earth
between two mirrors \cite{Jenke1}.

\section{Numerical solution of the problem}
\label{sec:cham-num}

In this section, we propose a numerical solution of Eq.(\ref{eq:34})
for a chameleon field in the spatial region $\textstyle z^2 \le
\frac{d^2}{4}$ between two mirrors with a density $\rho_m = 2.51\,{\rm
  g/cm^3}$ separated by the distance $d = 25.5\,{\rm \mu m}$
\cite{Jenke1}. First of all we would like note that due to a symmetry
of the spatial region a chameleon field as a scalar field is symmetric
with respect to a transformation $z \longleftrightarrow - z$,
i.e. $\phi(z ) = \phi(-z)$. For a numerical solution we transcribe
Eq.(\ref{eq:34}) into the form
\beq\label{eq:81}
 \sqrt{\frac{n}{2(n + 1)}}\,\frac{|z|}{\lambda_0} =
\int^{t(\phi)}_0\frac{dt}{\displaystyle (1 - t^2)^{\textstyle \frac{n
      - 2}{2 n}}},
\eeq
where $1/\lambda_0 = \Lambda \sqrt{n(n+1)}(\Lambda/\phi_0)^{\frac{n +
    2}{2}}$ , $\phi_0$ is given by Eq.(\ref{eq:43}) and $|z|$ is the
absolute value of $z$. As a result we define a chameleon field as a
function of $|z|$, i.e. $\phi(z) = f(|z|)$. Such a non--analytical
dependence of a chameleon field on a spatial variable does not prevent
it to satisfy the equation of motion Eq.(\ref{eq:28}). Indeed, the
first derivative of a chameleon field with respect to $z$ is equal to
\beq\label{eq:82}
 \frac{d\phi(z)}{dz} =
\varepsilon(z)\,\frac{df(|z|)}{d|z|},
\eeq
where $\varepsilon(z)$ is a sign function, defined by $d|z|/dz =
\varepsilon(z) = \theta(z) - \theta(-z)$ and $\theta(\pm z)$ are the
Heaviside functions \cite{GF64}. For the second derivative we obtain
the following expression
\begin{eqnarray}\label{eq:83}
\hspace{-0.3in}&&\frac{d^2\phi(z)}{dz^2} =
2 \delta(z)\,\frac{df(|z|)}{d|z|} + \varepsilon^2(z)\,\frac{d^2f(|z|)}{d|z|^2} = \nonumber\\
\hspace{-0.3in}&&= 2 \delta(z)\,\frac{df(|z|)}{d|z|}\Big|_{z = 0} +
\varepsilon^2(z)\,\frac{d^2f(|z|)}{d|z|^2} =
\frac{d^2f(|z|)}{d|z|^2},
\end{eqnarray}
where we have used that $d\varepsilon(z)/dz = 2 \delta(z)$ and
$\varepsilon^2(z) = 1$ \cite{GF64}. Thus, due to a continuous
derivative of a chameleon field at $z = 0$ a dependence of a chameleon
field on the absolute value of $z$ does not introduce an additional
terms violating the equation of motion Eq.(\ref{eq:28}).

For $n = 1$ the integral Eq.(\ref{eq:81}) can be calculated
exactly. We obtain
\beq\label{eq:84}
\frac{|z|}{\lambda_0} = t\sqrt{1 - t^2} +
       {\rm arcsin} t.
\eeq
This gives a chameleon field as a function of $t(|z|/\lambda_0)$
\beq\label{eq:85}
\phi(z) = \phi_0\Big(1 - t^2\Big(\frac{|z|}{\lambda_0}\Big)\Big).
\eeq
In Fig.\,1 we show the profiles of a chameleon field, obtained in the
spatial region $\textstyle z^2 \le \frac{d^2}{4}$ in the strong
coupling limit by a numerical solution of Eq.(\ref{eq:81}) for the
experimental mirror density $\rho_m = 2.51\,{\rm g/cm^3}$ and $d =
25.5\,{\rm \mu m}$ and $n \in [1, 10]$. The lower red line gives the
solution for $n = 1$, the green line corresponds to the solution for
$n = 2$ and so on.
\begin{figure}
\centering
\includegraphics[height=0.18\textheight]{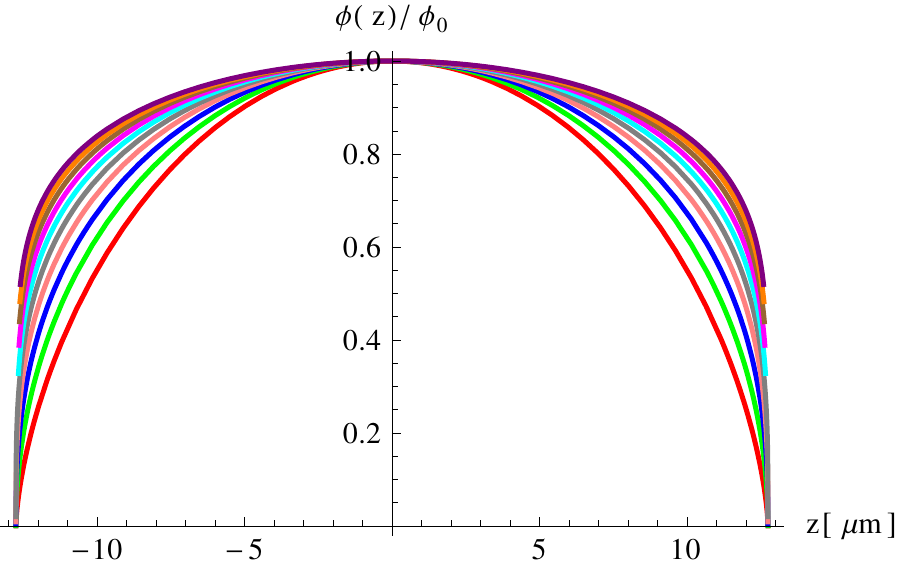}
\caption{The profiles of a chameleon field, calculated in the strong
  coupling limit as the solutions of Eq.(\ref{eq:81}) in the spatial
  region $\textstyle z^2 \le \frac{d^2}{4}$ and $n \in [1, 10]$. }
\end{figure}
 In Fig.\,2 we give the profiles of a chameleon field in the $3D$
 picture.
\begin{figure}
\centering
\includegraphics[height=0.35\textheight]{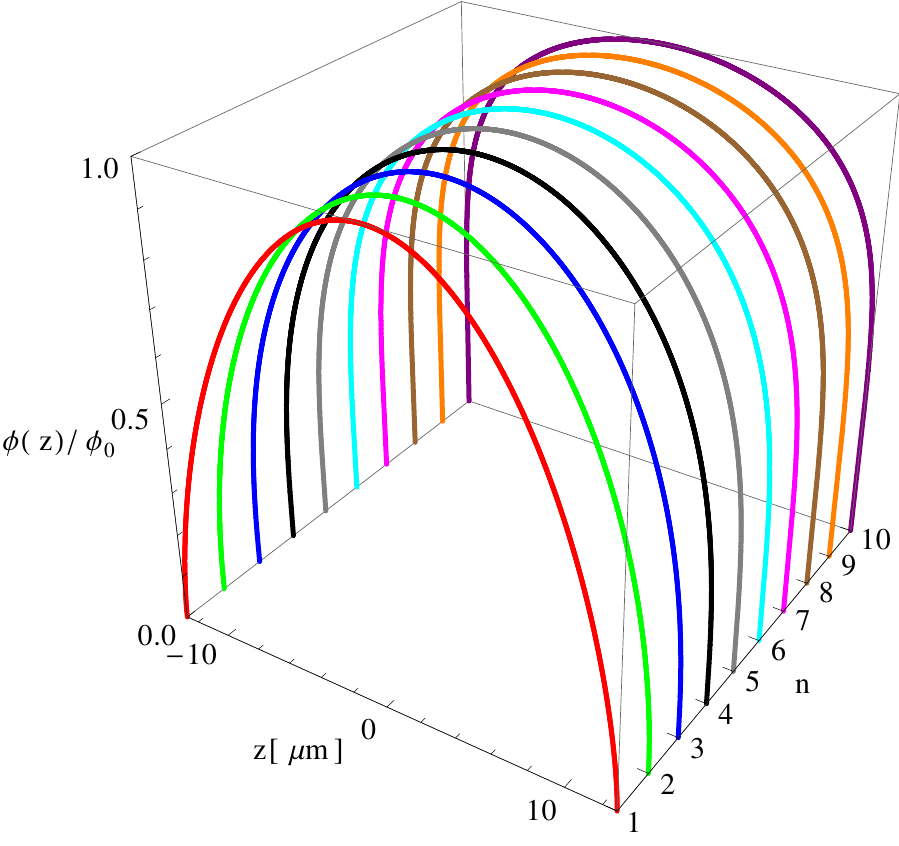}
\caption{The profiles of a chameleon field, calculated in the strong
  coupling limit as the solutions of Eq.(\ref{eq:81}) in the spatial
  region $\textstyle z^2 \le \frac{d^2}{4}$ and $n \in [1, 10]$. }
\end{figure}
In Fig.\,3 wWe compare the numerical solutions of Eq.(\ref{eq:81})
with the approximate solution Eq.(\ref{eq:73}), which we represent
in the more convenient form
\beq\label{eq:86}
\phi(z) = \phi_0\Big(\frac{n}{2(n +
  1)}\frac{1}{\lambda^2_0}\Big(\frac{d^2}{4} -
z^2\Big)\Big)^{\textstyle \frac{1}{n}},
\eeq
where $1/\lambda_0 = \Lambda \sqrt{n(n+1)}(\Lambda/\phi_0)^{\frac{n +
    2}{2}}$ but $\phi_0$ is given by Eq.(\ref{eq:66}).
\begin{figure}
\centering
\includegraphics[height=0.90\textwidth]{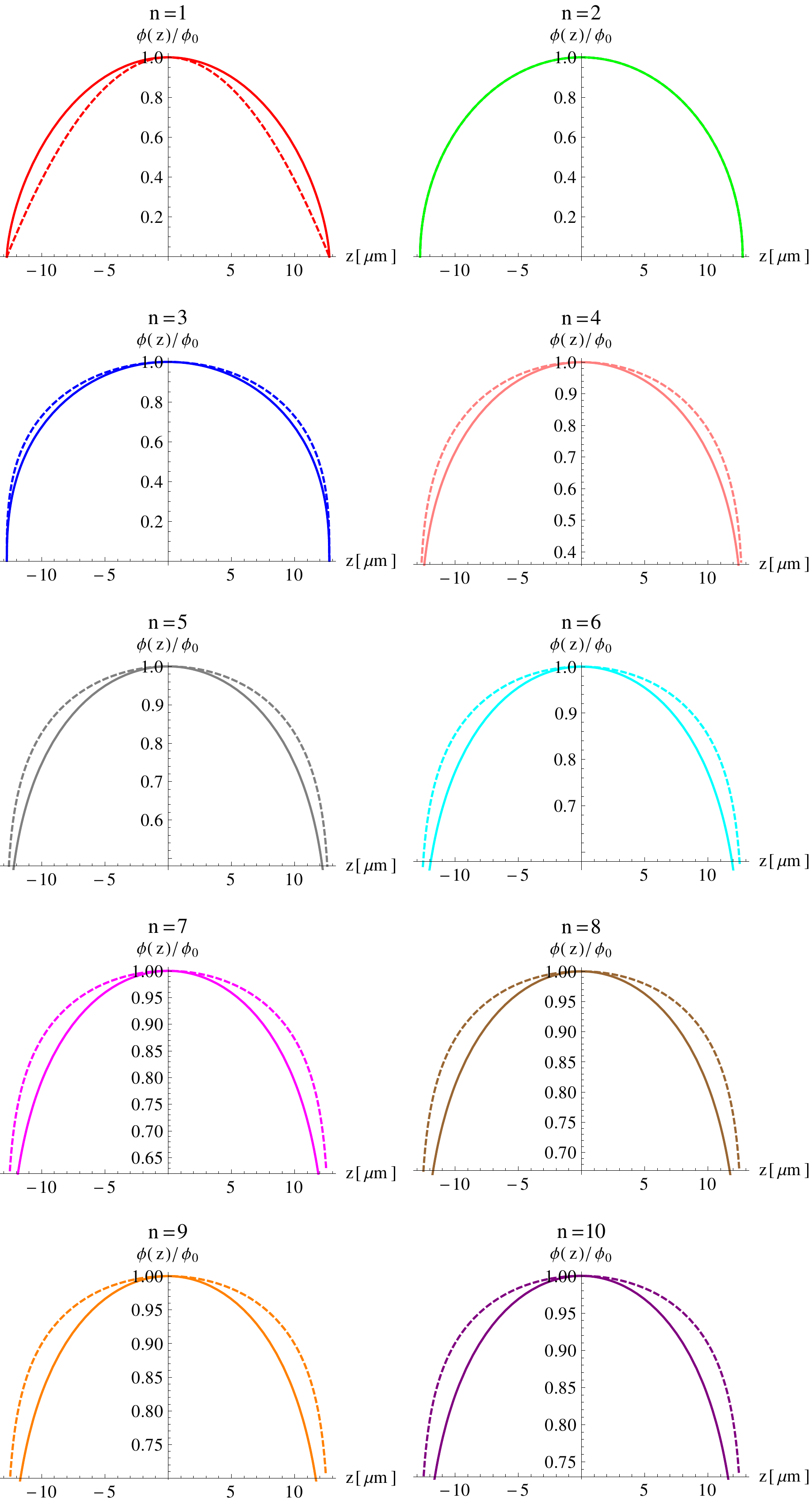}
\caption{The profiles (solid lines) of a chameleon field $\phi(z)/\phi(0)$ in the
  spatial region $\textstyle z^2 \le \frac{d^2}{4}$ as the solutions
  of Eq.(\ref{eq:81}), calculated numerically in the strong coupling
  limit, in comparison with the approximate solution Eq.(\ref{eq:86}
  (dashed lines) for $n \in [1, 10]$. }
\end{figure}
\begin{figure}
\centering
\includegraphics[height=0.70\textheight]{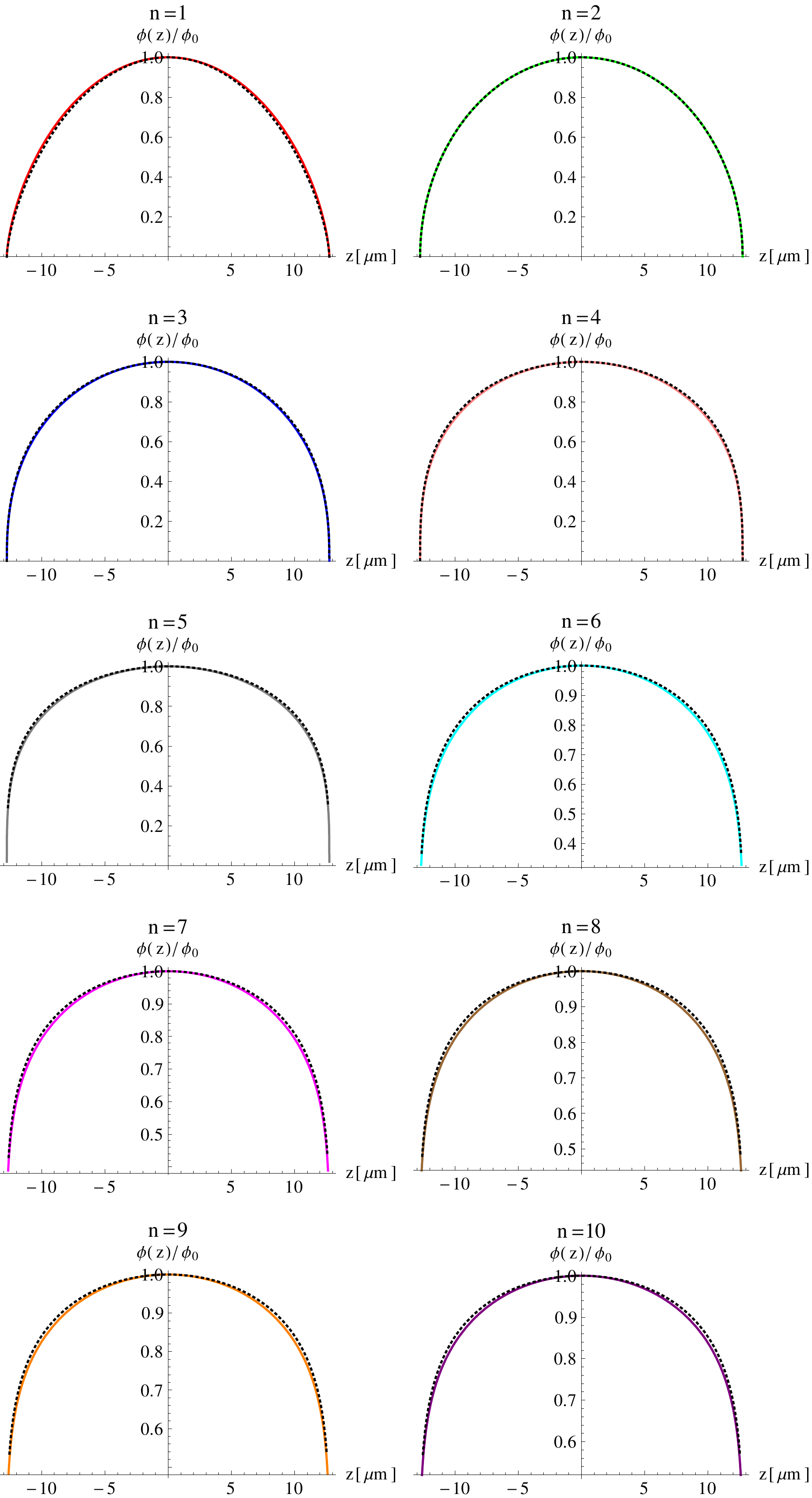}
\caption{The analytical fit of the numerical solutions of the
  non--linear equations of motion of a chameleon field $\phi(z)/\phi(0)$ in the spatial
  region $\textstyle z^2 \le \frac{d^2}{4}$, obtained in the strong
  coupling limit for $n \in [1, 10]$. }
\end{figure}
The dashed lines describe the approximate solutions, whereas the solid
line correspond to the exact numerical solutions of
Eq.(\ref{eq:81}). One can see that for $n = 2$ the dashed and solid
lines coincide, since for $n = 2$ the problem admits the analytical
solution in the elementary functions. Such an agreement confirms the
correctness all numerical solutions, obtained for $n \in [1,10]$.
Starting with $n \ge 3$ the approximate solution is larger compared
with the exact numerical solutions.

Following such a behaviour of the approximate solutions and keeping in
mind that the exact solutions of a chameleon field taken in the
vicinity of a mirror $z \simeq \mp \frac{d}{2}$ should reproduce the
solutions Eq.(\ref{eq:22}), for the fit of the exact numerical
solutions of a chameleon field
we propose the function
\begin{align}
\phi(z) &= \phi_0 \Big(1 - \frac{4z^2}{d^2}\Big)^{\textstyle \frac{2}{n
      + 2}} \label{eq:87}\\ &= \Lambda \Big(\frac{n +
    2}{\sqrt{2}}\,\frac{\Lambda}{d}\Big(\frac{d^2}{4} - z^2\Big)\Big)^{\textstyle
    \frac{2}{n + 2}}
\label{eq:87b},
\end{align}
where $\phi_0$ is equal to
\beq\label{eq:87c} \phi_0 =
\Lambda\,\Big(\frac{n + 2}{4\sqrt{2}}\,\Lambda d\Big)^{\textstyle
  \frac{2}{n + 2}},
\eeq
The results of the fit are shown in Fig.\,4. One may see that
Eq.(\ref{eq:87}) fits well the numerical solutions of the non--linear
equation of motion of a chameleon field and can be used for the
analytical description of a chameleon field in the spatial region
$\textstyle z^2 \le \frac{d^2}{4}$ between two mirrors with a density
$\rho_m = 2.51\,{\rm g/cm^3}$ in the strong coupling limit $\beta \ge
10^5$ and an arbitrary $n$. For $n > 2$ the parameter $\phi_0$, given
by Eq.(\ref{eq:87c}) reproduces the exact $\phi_0$, defined by
Eq.(\ref{eq:43}), with an accuracy better than $3.7\,\%$.

\section{Conclusion}
\label{sec:conclusion}

We have analyzed the solutions for a chameleon field, localized
between two mirrors in the spatial region $\textstyle z^2 \le
\frac{d^2}{4}$, and its influence on the transition frequencies of the
quantum gravitational states of ultra-cold neutron, bouncing between
two mirrors in the gravitational field of the Earth.

For the power $ n = 2$ of the potential of a chameleon field we have
found the exact analytical solutions of non--linear equations of
motion. We have shown that for the experimental density of mirrors $\rho_m
= 2.51\,{\rm g/cm^3}$ and a relative distance between mirrors $d =
25.5\,{\rm \mu m}$ \cite{Jenke1} the obtained solutions obey the
strong coupling regime with $\beta \ge 10^5$. In such a regime a
chameleon field does not depend on the coupling constant. As a result
the contribution of a chameleon field to the transition frequencies of
ultra-cold neutrons, bouncing in the gravitational field of the Earth
between two mirrors, is proportional to the coupling constant $\beta$.

In the strong coupling limit $\beta > 10^5$ we have
  obtained the approximate solutions of non--linear and linearized
  equations of motion for a chameleon field and compared them with the
  exact numerical solutions of non--linear equations of motion and the
  analytical fit of the exact numerical solutions of non--linear
  equations of motion. The analytical fit of the exact numerical
  solutions is taken in the form reproducing in the vicinity of a
  mirror at $z \simeq \mp \frac{d}{2}$ the solutions Eq.(\ref{eq:22}),
  coinciding with the solutions obtained in \cite{Brax1}. For $n > 2$
  the accuracy of the analytical fit of the exact numerical solutions
  is better than $3.82\,\%$.

Eq.~\ref{eq:87} may be used, for example, to calculate bounds on the coupling constant~$\beta$
by comparing the transition frequency~$\omega_{ab}$ with its theoretical expectation~$\omega_{ab}^{\rm theo}$:
\beq
\omega_{ab} - \omega_{ab}^{\rm theo} =
\beta \frac{m}{M} \left( \langle a | \phi(z) | a \rangle - \langle b | \phi(z) | b \rangle\right).
\label{eq:90}
\eeq
Such resonant transitions between quantum states of a neutron in the
gravity potential have been measured by the qBounce Collaboration
\cite{Jenke1}.  So far a chameleon field has been calculated in the
infinitely large spatial region above a mirror, whereas the ultra-cold
neutrons bounce in the gravitational field of the Earth between two
mirrors \cite{Jenke1} with a relative distance $d$.

Finally we would like to mention that the non--linear
Eq.(\ref{eq:27}) with $C = 0$ admits non--linear solutions with a
discontinuous derivative of a chameleon field. Such solutions are
equal to the solutions Eq.(\ref{eq:20}) with a replacement $z \to
|z|$. The first derivative of these solutions is proportional to the
sign function $\varepsilon(z)$ \cite{GF64}. Since $\varepsilon^2(z) =
1$ \cite{GF64}, the second derivative satisfies Eq.(\ref{eq:28})
with an additional term $-4\pi\,\sigma_d\,\delta(z)$, where a Dirac
$\delta$--function $\delta(z)$ appears as a derivative of the sign
function $\varepsilon'(z) = 2 \delta(z)$ \cite{GF64}. Then,
$4\pi\,\sigma_d$ is defined by a jump of the first derivative at $z =
0$, where $\sigma_d$ has a dimension of surface density of scalar
particle. The factor $4\pi$ is introduced by analogy with
electrostatic \cite{Jackson}. An analogy between the chameleon field
theory in the thin--shell regime and the electrostatic has been drawn
in \cite{ES1,ES2}.

A deviation from Eq.(\ref{eq:28}) by the term $-4\pi \sigma_d
\delta(z)$ might imply that the Hamilton density of a chameleon field
in the spatial region $\textstyle z^2 \le \frac{d^2}{4}$ should be
defined as follows
\beq\label{eq:89}
{\cal H}(z) =
\frac{1}{2}\,\Big(\frac{d\phi(z)}{dz}\Big)^2 + V(\phi(z)) - 4\pi
\sigma \delta(z) \phi(z).
\eeq
The parameter $\sigma$ can be unambiguously determined due to a
self--interaction of a chameleon field. Indeed, the Hamilton density
Eq.(\ref{eq:89}) defines the equation of motion Eq.(\ref{eq:28})
with the additional term $-4\pi \sigma \delta(z)$. Solving this
equation for the regions $\textstyle -\frac{d}{2} \le z < 0$ and $0 <
z \le + \frac{d}{2}$, respectively, we arrive at the solution
Eq.(\ref{eq:20}) with a replacement $z \to |z|$. This gives $\sigma
= \sigma_d$. When the distance between two mirrors tends to infinity,
the surface density $\sigma_d$ vanishes and we arrive at the solutions
for a chameleon field above (below) one mirror.

Of course, we do not stand for the reality of such solutions due to a
necessity to introduce an additional term $-4\pi \sigma\,
\delta(z)\,\phi$ to the Hamilton density, an influence of which on the
properties of a chameleon field is not clear. The solutions with a
discontinuous derivative might be accepted as an artifact of the
solutions of Eq.(\ref{eq:27}).

\begin{acknowledgments}
We thank M. Faber and G. Pignol for fruitful discussions.
We gratefully acknowledge support from the Austrian \textit{Fonds zur
  F\"orderung der Wissenschaftlichen Forschung} (FWF) under the
contract I862-N20 and the \textit{Deutsche Forschungsgemeinschaft} (DFG)
as part of the priority programme SPP~1491 \textit{Precision experiments
  in particle and astroparticle physics with cold and ultra-cold
  neutrons}.
\end{acknowledgments}

\end{document}